\documentclass[12pt]{article}

\usepackage{amsmath}
\usepackage{amssymb}
\usepackage{amsfonts}
\usepackage{eucal}
\usepackage{wasysym}

\newcommand{\bse}{\begin{subequations}}
\newcommand{\ese}{\end{subequations}}
\newcommand{\be}{\begin{equation}}
\newcommand{\bea}{\begin{eqnarray}}
\newcommand{\eea}{\end{eqnarray}}
\newcommand{\ba}{\begin{array}}
\newcommand{\ea}{\end{array}}
\newcommand{\ee}{\end{equation}}

\makeatletter
\@addtoreset{equation}{section}

\makeatother
\expandafter\ifx\csname mathbbm\endcsname\relax

\else

\fi
\def\cN{{\cal N}}
\def\cO{{\cal O}}
\def\Tr{{\rm {Tr}}}

\def\lp{l_{\rm Pl}}
\def\hmu{\hat{\mu}}
\def\hnu{\hat{\nu}}

\def\pl{plane-wave}

\def\lc{light-cone}
\def\Ham{Hamiltonian}
\def\bg{background}
\def\holo{holographic}

\def\rep{representation}

\def\sugra{supergravity}

\def\super{$PSU(2|2)\times PSU(2|2)\times U(1)$ }

\begin{document}   
\baselineskip 18pt

\begin{titlepage}
\hfill
\vbox{
    \halign{#\hfil         \cr
           SU-ITP-04/27\cr
           hep-th/0406214 \cr
           } 
      }  
\vspace*{6mm}
\begin{center}
{\Large {\bf Tiny Graviton Matrix Theory:\\
DLCQ of IIB Plane-Wave  String Theory, A Conjecture}}
\vspace*{5mm}
\vspace*{1mm}
                                                                                                         
{\bf 
M. M. Sheikh-Jabbari}
                                                                                                         
\vspace*{0.4cm}
{\it {Department of Physics, Stanford University\\
382 via Pueblo Mall, Stanford CA 94305-4060, USA}}\\
{E-mail:{\tt jabbari@itp.stanford.edu}}

\vspace*{1cm}
\end{center}
\begin{center}
{\bf\large Abstract}
\end{center}

We conjecture that the discrete light-cone quantization (DLCQ) of strings on the maximally 
supersymmetric type IIB plane-wave background in the sector with $J$ units of light-cone 
momentum is a supersymmetric $0+1$ dimensional $U(J)$ gauge theory (quantum mechanics) with 
\super superalgebra. The conjectured \Ham\ for the plane-wave matrix (string) theory, the tiny 
graviton matrix theory, is the quantized (regularized) three brane action on the same background.
We present some pieces of evidence for this conjecture through analysis of the \Ham , its 
vacua, spectrum and  coupling constant. Moreover, we discuss an extension of our conjecture to 
the DLCQ of type IIB strings on $AdS_5\times S^5$ geometry. 
\end{titlepage}
\section{Introduction}

According to Banks-Fischler-Susskind-Shenker (BFSS) conjecture  
discrete light-cone quantization (DLCQ) of M-theory  
in the sector with $N$ units of light-cone momentum is described by a 
supersymmetric quantum mechanics, a $0+1$ dimensional super Yang-Mills $U(N)$ gauge theory
\cite{BFSS}. The BFSS Hamiltonian is basically describing dynamics of $N$ D0-branes.
D0-branes from 11  dimensional viewpoint are gravity waves (gravitons) hence, according 
to BFSS proposal DLCQ of M-theory is described by $N$ gravitons of the same theory. 
This conjecture so far has passed many crucial tests, for a review and more detailed discussions 
see \cite{Taylor}. The original BFSS Hamiltonian was proposed to describe DLCQ of M-theory on 
the flat space background. This conjecture, however, was generalized to weakly curved 
backgrounds
\cite{Wati-Mark} and recently, by Berenstein-Maldacena-Nastase (BMN), to M-theory on 
the 
maximally supersymmetric 11 dimensional plane-wave background \cite{BMN}.  

The BFSS Matrix model is based on the D0-brane dynamics. The same action can also be obtained
from the regularized (``quantized'') M2-brane action in the 11 dimensional flat space
background, once the light-cone gauge is fixed \cite{Hoppe}. The same idea, i.e. quantization
of M2-branes
in the 11 dimensional plane-wave background, was shown to lead to BMN Matrix model \cite{DSV1}.
The BMN Matrix model has some specific features which makes it more tractable than the
BFSS case: there are no flat directions and hence (at least for finite $N$) BMN Matrix model 
has a large set of normalizable
supersymmetric vacua all of them have interpretation in terms of spherical M2-brane giant
gravitons in the 11 dimensional plane-wave background \cite{BMN, DSV1}. Giant gravitons, are
spherically shaped D or M -branes, which are following light-like geodesics and the spatial part
of their worldvolume is a sphere, blown up due to the existence of non-zero form flux in the
background \cite{MST}. Intuitively, one may interpret the BMN Matrix model as a theory of $N$
spherical M2-branes of very small size each of which carrying one unit of the light-cone 
momentum. For the reasons which will become apparent in the
next section, these objects will be called (eleven dimensional) ``tiny'' gravitons, i.e. DLCQ of 
M-theory  on the 11 dimensional plane-wave background is a theory of $N$ tiny (membrane) 
gravitons.

In this paper, inspired by the BFSS idea, and in light of the above description of BMN matrix 
model, we present a conjecture for the DLCQ of type IIB string theory on the maximally 
supersymmetric ten dimensional  plane-wave background. This background, which hereafter would be 
referred to as ``the'' plane-wave to distinguish it from other plane-wave geometries, as a 
solution of type IIb supergravity is given by
\begin{subequations}\label{background}
\begin{align}
  ds^2  =  -2 dX^+ dX^- & -\mu^2(X^i X^i + X^a X^a) {(dX^+)}^2 + dX^i dX^i+ dX^a dX^a \, ,\\
  F_{+ijkl} &= \frac{4}{g_s} \mu\ \epsilon_{ijkl} \, ,\ \ \ \ \ \
  F_{+abcd}=\frac{4}{g_s} \mu \ \epsilon_{abcd} \ ,
\end{align}
\end{subequations}
where $i,a=1,2,3,4$.
The above (self-dual) fiveform field is the field strength of the fourform $C_4\ (F=dC_4)$ 
\be\label{fourform}
  C_{+ijk} = -\frac{\mu}{g_s} \epsilon_{ijkl} X^l \, , \ \ \ \
  C_{+a b c} = -\frac{\mu}{g_s} \epsilon_{a b c d} X^{d} \, .
\ee
Here we follow conventions and notations of \cite{review}.
This background has a globally defined light-like Killing vector $\partial/\partial X^-$.
String theory $\sigma$-model on this background, in the light-cone gauge, is shown to be 
solvable \cite{Metsaev}. And again according to BMN \cite{BMN}, as a specific case (or extension 
of) the usual AdS/CFT duality \cite{Magoo}, type IIB string theory on the plane-wave \bg\ has 
been conjectured to be dual to a specific 
subsector, known as BMN sector, of ${\cal N}=4,\ D=4$ $U(N)$ SYM gauge theory.
The BMN sector consists of ``almost'' chiral primary operators with large R-charge $J$, 
$J\sim \sqrt{N}$, and by almost chiral-primary we mean operators of conformal dimension $\Delta$
 and R-charge $J$ where $\Delta-J\ll J$. For more detailed discussion on the plane-wave/SYM 
duality see \cite{review, Plefka-lectures}.

The plane-wave \eqref{background} has a one dimensional light-like boundary \cite{BN} and one 
might wonder whether besides the BMN description, similarly to the AdS/CFT case \cite{SW}, we 
have a \holo\ description for strings on the plane-wave \bg .\footnote{The plane-wave geometry 
\eqref{background} can be obtained from $AdS_5\times S^5$ solution upon taking the Penrose limit 
\cite{BMN}. However, the causal structure, the Penrose diagram and hence the boundary of the 
plane-wave are different than $AdS_5\times S^5$.} 

This \holo\ description, if it exists, should then be a $0+1$ dimensional theory, presumably a 
gauge theory. The goal of this paper is to propose a possible candidate for such a \holo\ 
formulation. For some previous attempts in this direction see \cite{BN, Gopakumar}. Noting that 
the boundary of the \pl\ is light-like, this \holo\ description for the sector of a given \lc\ 
momentum is then expected to provide us with the DLCQ of strings on the type IIB plane-wave \bg 
.

The paper is organized as follows. In section \ref{proposal}, after introducing the ``tiny'' 
gravitons and some suggestive arguments, we present the statement of our conjecture and the 
proposed \lc\ \Ham\ for the DLCQ of strings on the plane-wave \bg . In section 
\ref{Gauss-law-section}
we present the condition (the Gauss law constraint) that all physical states of the Matrix 
model should satisfy. In section 
\ref{String-limit} 
we discuss the limit under which we recover the type IIB string theory. In section 
\ref{Extension} we discuss the extension of our conjecture to the DLCQ of type IIB strings on 
$AdS_5\times S^5$ geometry. In section \ref{Evidence}, we present some pieces of evidence in 
support of our proposal. These include analysis of the vacua of the theory and the spectrum 
about these vacua. We close with conclusions and discussions. In the four Appendices we have 
gathered necessary technical points. In Appendix \ref{LCHam}, we present  the \lc\ \Ham\ for a 
3-brane in the \pl\ \bg. In Appendix \ref{Quantum-Nambu}, we  briefly review Nambu brackets 
and their quantization. In particular we present a novel solution to the problem of 
quantization of Nambu odd brackets.
In Appendix \ref{SUSY-algebra}, we have presented the superalgebra of the 
\pl\ \bg\ and its representation in terms of  matrices of our tiny graviton Matrix theory.
Appendix \ref{S3F} contains a brief review on fuzzy spheres, and in particular fuzzy three 
sphere. In this Appendix we also present a novel and unified way for fuzzifying generic $d$ 
spheres.

\section{The proposal}\label{proposal}

Our proposal is that the DLCQ of strings on the ten dimensional plane-wave \bg\ in the sector
with $J$ units of light-cone  momentum, similarly to the BMN Matrix model case ({\it cf.} 
discussions of  Introduction), is described by the dynamics of $J$ ``tiny'' gravitons, which 
are very small size 3-branes, on this 
\bg .  In order to motivate and explain our proposal let us recall 
the giant gravitons in $AdS_5\times S^5$ geometry. Following \cite{MST},  consider a 
3-brane moving along a light-like geodesic in the $S^5$ with angular momentum $J$. 
In two of the three minimum energy, half BPS configurations (corresponding to the giant 
gravitons 
grown in $S^5$ or in $AdS_5$ \cite{AdS-giants}) 3-brane worldvolume is a three sphere  of radius 
\be\label{giant-radius-AdS-units}
\left(\frac{R_{giant}}{R_{AdS}}\right)^2=\frac{J}{N}, 
\ee
where $N$ is the number of fiveform fluxes on $S^5$.
Next recall that in the $AdS_5\times S^5$ geometry,
\be\label{AdSR-string-units}
\left(\frac{R_{AdS}}{l_s}\right)^4=g_s N
\ee
Noting that ten dimensional Planck length $\lp$ and the string scale $l_s$ are related as
$\lp^4=l_s^4 g_s$, we have 
\be\label{AdSR-planck-units}
\left(\frac{R_{AdS}}{\lp}\right)^4=N.
\ee

Therefore, the size of the smallest giant graviton, which will henceforth be called a {\it 
tiny} graviton, 
i.e. the case in which the 3-brane is carrying one unit of angular momentum $J=1$, is given 
by\footnote{ One should note that the notion of size for giant gravitons we have been using so 
far, like the giant graviton themselves, is a classical one. Indeed,  the main idea and 
subject of the paper is how to quantize them.}
\be\label{tiny-planck-AdS}
R_{tiny}=\frac{\lp^2}{R_{AdS}}.
\ee
This is a very remarkable result: For the cases where we can trust the \sugra\ description, 
$R_{AdS}\gg \lp$, $R_{tiny}\ll \lp$ and hence we would expect the fundamental description of the 
theory should come from the {\it tiny} gravitons.\footnote{
These results may be repeated for  M$p$ tiny gravitons, $p=2,5$ in $AdS_{4,7}\times S^{7,4}$ 
or the corresponding 11 dimensional plane-wave background. The analog of \eqref{tiny-planck-AdS}
for $AdS_{q}\times S^{D-q}$ space (D=11) is 
\be\label{M-tiny-size}
R^{p-1}_{tiny} R^{q-p}_{AdS}=\lp^{q-1}.
\ee
Note that now $\lp$ is the 11 dimensional Planck length. It is amusing that for $p=3$ 
\eqref{M-tiny-size} reproduces \eqref{tiny-planck-AdS}.}
Eq.\eqref{tiny-planck-AdS} also justifies why we have called these objects tiny gravitons.

The tiny gravitons, being 3-branes of small size, should show a brane-like behaviour. Namely,
we expect that when $J$ number of them sit on top of each other we should have a $U(J)$ gauge 
theory description. This result about the enhancement of the gauge symmetry, although 
non-trivial, 
has been supported by studying BPS open string-type states in the giant graviton worldvolume 
theory \cite{Hedgehog}. In this respect tiny gravitons are similar to D0-branes of BFSS 
and may be used to obtain a (non-perturbative) DLCQ formulation for string theory on 
$AdS_5\times S^5$ or the corresponding plane-wave. 

To obtain the action for $J$ tiny gravitons, we follow the logic of \cite{Hoppe} where the 
corresponding Matrix model is obtained as a regularized (quantized) version of M2-brane \lc\ 
\Ham , but now for 3-branes.
In other words, we conjecture that DLCQ of type IIB strings on the \pl\ \bg\ \eqref{background}
is nothing but a quantized 3-brane theory.

The 3-brane \lc\ \Ham\ for the plane-wave \bg \ is given in Appendix \ref{LCHam}, in 
eqs.~\eqref{Hlc-bosonic} and \eqref{Hlc-fermionic}. Since we are looking for a DLCQ 
description, we need to compactify $X^-$ on a circle of radius $R_-$:
\be\label{R-radius}
X^-\equiv X^- +2\pi R_-\ .
\ee 
This  leads to the quantization of the light-cone momentum $p^+$,
\be\label{p+J}
p^+=\frac{J}{R_-} \ .
\ee
In our conventions $\mu$ and  $R_-$ have dimension of energy, $p^+$ has dimension 
length and hence $\mu p^+$ is dimensionless.\footnote{
Note that we have set $\alpha'=l_s^2=1$. Recovering the powers of $\alpha'$, $\mu 
p^+\alpha'$ is the dimensionless, physical light-cone momentum.}

To quantize the 3-brane theory, following \cite{Taylor, Hoppe, DSV1}, we prescribe:
\newline
Replace $X^I(\sigma), P(\sigma)_I, \ 
\psi_{\alpha\beta}(\sigma),\  \psi_{\alpha\beta}(\sigma)$ with $J\times J$ matrices, i.e.
\be\label{func-to-mat}
\begin{split}
X^I(\sigma) & \longleftrightarrow X^I_{Mat} \cr   
P^I(\sigma) & \longleftrightarrow J\ \Pi^I_{Mat}\\  
\psi(\sigma) & \longleftrightarrow \sqrt{J}\ \psi_{Mat}\ ,
\end{split}
\ee
together with
\be\label{int-to-trace}
\frac{1}{p^+}\int d^3\sigma \ * \longleftrightarrow  R_- \Tr\ *
\ee
\be\label{Nambu-to-commutator}
\{ F, G, K\} \longleftrightarrow J\ [F, G, K, {\cal L}_5] .
\ee
In the above, and in what follows, we will use $I, J$ indices, $I, J=1,2,\cdots, 8$ to denote
the eight transverse directions, i.e. $X^I=(X^i, X^a)$.
For  the definition of quantized Nambu bracket see Appendix \ref{Quantum-Nambu} and for the 
definition of the $J\times J$ matrix ${\cal L}_5$ see Appendix \ref{S3F}, eqs.\eqref{S3F-JJ}, 
\eqref{size-rep}. 
After the above replacements we obtain the tiny graviton Matrix theory \Ham\ which  we 
conjecture to give the non-perturbative description of strings on the \pl\ in the sector 
with $J$ units of \lc\ momentum. Explicitly, the statement of our conjecture is:
\vspace{3mm}

{\it The theory of $J$ tiny  gravitons, which is a $U(J)$ supersymmetric quantum mechanics 
with the \super symmetry, is the Matrix theory describing the DLCQ of string on 
plane-waves with light-cone momentum $p^+=J/R_-$, $R_-$ is the light-like compactification 
radius. The \Ham \ of this Matrix model is:} 
\be\label{Matrix-model-Ham}
\begin{split}
{\bf H}= R_-\ \Tr&\Biggl[ \frac{1}{2}\Pi_I^2+ \frac{1}{2}\left(\frac{\mu}{R_-}\right)^2 
X_I^2 +\frac{1}{2\cdot 3! g_s^2} [ X^I , X^J , X^K, {\cal L}_5][ X^I , X^J , X^K, {\cal L}_5]\cr
&-\frac{\mu}{3!R_- g_s}\left(
\epsilon^{i j k l} X^i [X^j, X^k, X^l, {\cal L}_5]+ \epsilon^{a b c d} X^a [ X^b, X^c,
X^d , {\cal L}_5] \right)\cr
&+\left(\frac{\mu}{R_-}\right) \left(\psi^\dagger {}^{\alpha \beta} \psi_{\alpha \beta}-
\psi_{\dot\alpha \dot\beta}\psi^\dagger {}^{\dot\alpha \dot\beta}\right)\cr
&+\frac{2}{g_s}\left( \psi^\dagger {}^{\alpha \beta} (\sigma^{ij})_\alpha^{\:  \: \delta} 
  [ X^i, X^j, \psi_{\delta \beta}, {\cal L}_5] +
  \psi^\dagger {}^{\alpha \beta} (\sigma^{ab})_\alpha^{ \: \: \delta} \:
  [ X^a, X^b, \psi_{\delta \beta}, {\cal L}_5]\right) \cr
&-\frac{2}{g_s} \left(\psi_{\dot\delta \dot\beta}
(\sigma^{ij})_{\dot\alpha}^{ \: \: \dot\delta} \:
  [ X^i, X^j, \psi^\dagger {}^{\dot\alpha \dot\beta}, {\cal L}_5]+
\psi_{\dot\delta \dot\beta} 
(\sigma^{ab})_{\dot\alpha}^{\: \: \dot\delta} \:
  [ X^a, X^b, \psi^\dagger {}^{\dot\alpha \dot\beta}, {\cal L}_5]\right)\Biggr]\ .
\end{split} 
\ee
\vspace{2mm}

One of the advantages of our proposal (e.g. compared to that of \cite{Gopakumar}) is that
it explicitly exhibits the invariance under the  \super superalgebra, which is the superalgebra 
of 
the plane-wave background; see Appendix \ref{SUSY-algebra} for the  superalgebra and its 
representation in terms of the $J\times J$ matrices. The other advantage is that, similarly to 
BMN Matrix model \cite{BMN, DSV1}, there are no flat directions and the flat directions are 
lifted by the mass terms coming form the \bg\ \pl \ metric.  

The  $U(J)$ gauge symmetry of the above \Ham\ is in fact a discretized (quantized) form of
the spatial diffeomorphisms of the 3-brane. As is evident from the above construction (and 
the discussions of the Appendices) we expect in $J\to\infty$ limit to  recover the
diffeomorphisms.
In this respect, it is very similar to the usual 
BFSS Matrix model in which the gauge symmetry is the regularized form of the diffeomorphisms on 
the membrane worldvolume \cite{Hoppe}. 
The quantization of diffeomorphisms on a three dimensional surface (in our case a three 
sphere), unlike the two dimensional surfaces corresponding to membranes, with a desirable 
continuum limit is a formidable task. The fuzzy sphere $S^3_F$, however, provides us with a 
unique solution ({\it cf.} Appendix \ref{S3F}).
Moreover, the above proposal closely parallels a similar 
conjecture about M-theory on the 11 dimensional plane-wave \bg , the BMN Matrix model, where 
tiny membrane gravitons on the 11 dimensional plane-wave \bg\ are the fundamental 
objects of BMN Matrix theory \cite{DSV1}.

It is worth noting that in the  classical \Ham\ we started with, \eqref{Hlc-bosonic},  
we have not included the gauge field in the original Born-Infeld action and as such one may 
worry that we have not included all the necessary degrees of freedom to begin with.
As we will show in section \ref{X=JSpectrum},  these $U(1)$ gauge fields, and the 
corresponding photon states are indeed already accounted for and they appear in the   
spectrum of theory once we expand the theory about its vacua.

\subsection{Physical states}\label{Gauss-law-section}

The \Ham\ \eqref{Matrix-model-Ham} can be obtained from a $0+1$ dimensional $U(J)$ gauge theory 
Lagrangian, in the 
temporal gauge. Explicitly, the only component of the gauge field, $A_0$, has been set to zero. 
To ensure the $A_0=0$ gauge condition, all of our physical states must satisfy the Gauss law 
constraint arising from equations of motion of $A_0$. Similarly to the BFSS \cite{BFSS} and 
BMN \cite{DSV1} cases, these constraints, which consists of $J^2-1$ independent conditions are:
\be\label{Gauss-law}
\left(i[X^i, \Pi^i]+i[X^a, \Pi^a]+ 
2\psi^{\dagger\alpha\beta}\psi_{\alpha\beta}
+2\psi^{\dagger \dot\alpha\dot\beta}\psi_{\dot\alpha\dot\beta}\right)|\phi\rangle_{phys}= 0.
\ee
These constraints are the requirement of $SU(J)$ invariance of the physical states.

One may trace back the Gauss law constraint \eqref{Gauss-law} to the continuum 3-brane 
action. For this, recall that in the procedure of fixing the \lc\ gauge we need to impose 
constraints \eqref{level-matching}. (These constraints in the string theory language give rise 
to the level matching condition on the closed string states \cite{Polchinski}.) 
Upon quantization, and adding the fermionic terms, \eqref{level-matching} goes over to the 
Gauss law constraint \eqref{Gauss-law} which should be imposed on the physical states.

\subsection{String theory limit}\label{String-limit}

The \Ham\ \eqref{Matrix-model-Ham} is proposed to describe type IIB string theory on the \pl\ 
with 
compact $X^-$ direction. The ``string theory limit'' is then a limit where $R_-$ is taken to 
infinity, keeping $p^+$ fixed, i.e.
\be\label{string-theory-limit}
J, R_- \to \infty, \qquad \mu,\ p^+=J/R_-, g_s\ \ {\rm fixed}\ .
\ee
In fact one can show that in the above string theory limit one can rescale $X$'s such that 
$\mu, p^+$ only appear in the combination $\mu p^+$. Therefore the only parameters of the 
continuum theory are $\mu p^+$ and $g_s$. The \pl\ that we obtain 
after the Penrose limit, and hence the usual BMN double scaling limit, would appear 
in our model after taking the string theory limit.

One of the questions that often arise in the DLCQ descriptions of flat space (and in particular 
the BFSS Matrix model) is whether we can recover a covariant theory in the continuum limit. 
Here we briefly discuss this question posed for our tiny graviton Matrix theory. The first 
point we would like to remind the reader is an important difference between \pl\ \bg\ and the 
flat space. For concreteness let us compare the isometries of the ten dimensional \pl\ 
\eqref{background} and a ten dimensional flat space. The \pl\ metric has 30 isometries, 14 of 
which are $SO(4)\times SO(4)$ rotations and translations along $X^+, X^-$. The other 16 are 
forming a Heisenberg-type algebra, in particular we note that generator of translations along 
$X^-$, $p^+$, commutes with all the isometries and the light-cone boosts, $J^{+-},\ J^{I-}$ are 
all absent \cite{review}. This should be contrasted with the flat space Poincare group which is 
$10+45=55$ dimensional, including light-like boosts $J^{+-},\ J^{I-}$. Therefore, in the above 
formulation of the \pl\ string theory, which makes all the ``dynamical'' and 
``kinematical'' isometries manifest, the DLCQ description is already a complete one.
(For the definition of kinematical and dynamical isometries, which parallels the same 
terminology about supercharges see \cite{review}). In fact one can show that the kinematical 
isometries, similarly to the BMN case \cite{DSV1}, can be represented in terms of the $U(1)$ 
part (trace part) of $U(J)$ matrices.

\subsection{Extension of the proposal to DLCQ of strings on $AdS_5\times S^5$}\label{Extension}

In this section we extend our conjecture to beyond the Penrose limit, to the full $AdS_5\times 
S^5$. The proposal is that the \Ham\ for DLCQ of strings on $AdS_5\times S^5$ with $N$ units of 
RR flux on $S^5$, is the same as \eqref{Matrix-model-Ham} with $R_-=\sqrt{g_s N}$. In this 
case the gauge group can  be  as large as $U(N)$. The string theory limit 
\eqref{string-theory-limit} is then equivalent to large $N$ ($N\to \infty$) limit.

Since we are looking for a DLCQ description we need to 
consider light-like geodesics in $AdS_5\times S^5$. There are two kinds of them: those along 
the 
radial direction in the $AdS_5$ and the ones which are along a circle inside $S^5$. The former 
geodesics hit the boundary, moreover it is not possible to compactify the radial direction on a 
circle. So, for the purpose of DLCQ we are only left with the latter, geodesics in $S^5$. The 
objects which follow such geodesics are gravitons of various size, giant, normal or tiny 
gravitons. In fact sitting in the rest frame of the giant (or tiny) gravitons, from the 
viewpoint of the observer who uses the usual $AdS_5\times S^5$ global time as time 
coordinate, is 
like boosting to the infinite momentum frame, in which the strings and giants are made out of 
tiny graviton ``partons''.
Next, we note that our earlier discussion in the opening of this section and 
in particular \eqref{tiny-planck-AdS} holds for the $AdS_5\times S^5$. 

The above proposal, if correct, would shed light on the stringy exclusion principle \cite{MST}. 
It would also be very desirable to rederive or confirm the above proposal from the usual 
${\cal N}=4$ CFT description. We will comment more on this proposal in the discussion section, 
however, it needs a thorough and careful analysis. We hope to address this in a future work 
\cite{work-in-progress}.

\section{Evidence for the proposal}\label{Evidence}

In this section we present some evidence and arguments in support of our conjecture some of 
which we have already discussed. First, we note that by construction our $0+1$ dimensional 
gauge theory is  supersymmetric one with the \super superalgebra, which is the superalgebra of 
the \pl\ \bg. (We emphasize that this gauge theory is {\it not} a Yang-Mills theory). 
Furthermore, generically supersymmetric quantum mechanics with 16 supercharges 
are uniquely determined once we specify superalgebra and the gauge group. Hence, if the type 
IIB string on the \pl\ \bg\ admits a \holo\ description our \Ham\ is the first viable 
candidate. Indeed confirmation of the uniqueness of the $U(J)$ quantum mechanics with 
\super supersymmetry, if we believe in holography, can be regarded as a proof for our proposal 
\cite{work-in-progress}.
To see the second piece of evidence we work out the vacua and zero energy configurations.

\subsection{Vacua and zero energy solutions}\label{vacuum-solutions-subsection}

Inspired by the analysis of the continuum case \cite{Hedgehog}, we note that for the static 
($\Pi=0$), bosonic ($\psi=0$) configurations the \Ham\ takes the form
\be\label{vacuum-potential}
\begin{split}
V= R_-\ \Tr\Biggl[ 
&\frac{1}{2}\left(\frac{\mu}{R_-} X^l + \frac{1}{3! g_s}\epsilon^{ijkl} [ X^i , X^j , X^k, 
{\cal 
L}_5]\right)^2 \cr
+&\frac{1}{2}\left(\frac{\mu}{R_-} X^d+ \frac{1}{3! g_s} \epsilon^{abcd}[ X^a , X^b , X^c, 
{\cal 
L}_5]\right)^2\cr
+&\frac{1}{4g_s^2}\left(
[X^i, X^j, X^a, {\cal L}_5][ X^i, X^j,X^a , {\cal L}_5]+ 
[X^i, X^a, X^b, {\cal L}_5][ X^i, X^a,X^b , {\cal L}_5] 
\right)\Biggr].
\end{split}
\ee
Evidently, there are three types of zero energy vacuum configurations:
\begin{subequations}\label{vacua-equations}
\begin{align}
X^a&=0\ , \qquad   [ X^i , X^j , X^k, {\cal L}_5]= -\frac{\mu g_s}{R_-} \epsilon^{ijkl} X^l\\
X^i&=0\ , \qquad   [ X^a , X^b , X^c, {\cal L}_5]= -\frac{\mu g_s}{R_-}  \epsilon^{abcd} X^d\\
X^i&=X^a=0\ .
\end{align}
\end{subequations}
The physics of the first two solutions, (\ref{vacua-equations}a) and (\ref{vacua-equations}b), 
are very similar, because there is a $Z_2$ symmetry of the \bg\ \pl\ under which 
$X^i\leftrightarrow X^a$. Of course if we choose to expand the theory about either of these 
vacua, this $Z_2$ symmetry is spontaneously broken. Therefore, we only focus on solutions of
(\ref{vacua-equations}a) and (\ref{vacua-equations}c).

\vspace{2mm}
{\it  Solutions of (\ref{vacua-equations}a)}
\vspace{2mm}

Solutions of (\ref{vacua-equations}a) can be classified in terms of 
(irreducible or reducible) \rep s of $Spin(4)$ and in general they are of the form of 
concentric fuzzy three spheres of various radii ({\it cf.} Appendix \ref{S3F}). 
(This is very similar to the case of BMN Matrix theory studied in detail in \cite{DSV1}.)
The size of each three sphere is given by the fraction of the total light-cone momentum $J$ 
that it carries (where radius squared $\propto$ fraction of \lc\ momentum) and 
the sum of radii squared of these fuzzy spheres is $\frac{\mu g_s}{R_ -} J$. 

Here we only focus on the solution given by irreducible $Spin(4)$ \rep\ which corresponds to a 
single $S^3_F$. The reducible \rep s and their detailed analysis is postponed to  future works.
Noting \eqref{S3F-def} and \eqref{radius-Jn}, the radius of this $S^3_F$, in units of $l_s$, is
\be\label{single-giant-radius}
R^2=\frac{\mu g_s}{R_-} J = \mu p^+ g_s
\ee
with the fuzziness $l$:
\be\label{fuzziness-l}
l^2=\frac{\mu g_s}{R_-}l_s^2\ .
\ee
(Note that in \eqref{fuzziness-l} we have reintroduced factors of $l_s$.)
In the second equality of \eqref{single-giant-radius} we have used definition of $p^+$ 
\eqref{p+J}.
This irreducible solution, which will be denoted by $X=J$ vacuum, is indeed the giant graviton 
of \cite{MST, Hedgehog}. In other words, our tiny graviton Matrix theory contains giant 
gravitons as  zero energy configurations. In our picture a giant graviton of radius $R$ is a 
state in which $J$ tiny gravitons are blown up into a three sphere.

Here we would like to compare \eqref{single-giant-radius} to results of \cite{MST, Hedgehog}. 
As we see in both cases, i.e. \eqref{giant-radius-AdS-units} and \eqref{single-giant-radius}, 
$R^2$ is proportional to $J$. We would like to present this fact as the second evidence in 
support of our proposal. This point, which may seem a trivial result of our construction, is 
very remarkable. Note that \eqref{giant-radius-AdS-units} is coming from a physical condition, 
namely stability of the spherical brane (and in general for a $p$ sphere giant $p=2,3,\ 5$
$R^{p-1}\propto J$ \cite{MST}), whereas \eqref{single-giant-radius} is a result of 
$Spin(4)=SU(2)\times SU(2)$ group theory and its \rep s.\footnote{Of course a similar relation 
also holds for spherical M2-branes \cite{DSV1}, while is not true for M5 giants. We will 
comment more on this point in  section \ref{conclusion}.}
This result becomes less trivial noting that, unlike the membrane case, there is no 
unique prescription for quantization of diffeomorphisms of a three dimensional surface. In this 
regard \eqref{single-giant-radius} is a confirmation of the fact that our quantization 
proposal, namely replacing a three sphere giant with a fuzzy three sphere, is the right one. 

\vspace{2mm}
{\it The (\ref{vacua-equations}c) solution, $X=0$ vacuum}
\vspace{2mm}

The physical interpretation of (\ref{vacua-equations}c) solution, the $X=0$ vacuum, is quite 
non-trivial; as the second part of our proposal we conjecture that 

\vspace{2mm}
{\it In the string theory limit 
discussed in  section \ref{String-limit}, this vacuum quantum mechanically becomes the vacuum 
for strings 
on the \pl\ \bg , i.e. the BMN vacuum \cite{BMN}.  The fundamental type IIB closed strings, 
however,  appear as non-perturbative objects in this vacuum.}

As first evidence for the above proposal, we note that:
\newline
{\it i)} all of these vacua, reducible or irreducible are half BPS states. That is, they 
preserve all the dynamical supercharges. To see this recall the superalgebra 
\eqref{susy-algebra} and the 
fact that all of these vacua are zero energy solution with ${\bf J}_{ij}={\bf J}_{ab}=0$.
\newline
{\it ii)} In string theory limit \eqref{string-theory-limit}, among the (infinitely many) 
vacua,
reducible or irreducible, we only remain with $X=0$ and two irreducible, single giant $X=J$ 
vacua. In the BMN gauge theory, these states should correspond to chiral-primary operators 
\cite{BMN} and the only single particle states available  are either single string 
(BMN 
vacuum) or single giant graviton vacua.\footnote{ 
Some brief comments on the connection to the usual ${\cal N}=4, D=4$ gauge theory are in order:

Denote one of the three complex scalars in ${\cal N}=4, D=4,\ U(N)$ SYM theory by $Z$, the 
chiral-primary 
operators are then made 
out of gauge invariant products of $Z$'s.
Let us focus on chiral-primaries with R-charge $J$ (i.e. they are composed of $J$ number of 
$Z$'s.) The gauge invariant operators are then obtained by summing over all $U(N)$ indices. 
There are three different ways of constructing such gauge invariant operators: using trace, 
(sub)determinant \cite{Bala-Strassler} and Schur's polynomials \cite{Jevicki}. In the BMN 
limit, however, the latter two would essentially become identical. We then remain with the two 
following possibilities to expand a chiral-primary operator of R-charge $J$:
\begin{subequations}\label{chiral-primary-basis}
\begin{align}
{\rm Trace\ basis:}\qquad\ \Tr Z^J,\ \ :\Tr Z^{J_1}\Tr Z^{J-J_1}:,\ &\ \cdots ,\ \ :(\Tr 
Z)^J:\\
{\rm subdeterminant\ basis:}\qquad\ \cO_J\ ,\ \ :\cO_{J_1}\cO_{J-J_1}:,\ &\ \cdots ,\ \ :\cO_1^J: 
\end{align}
\end{subequations}
with
\be
\cO_k=\cN_k 
\epsilon_{i_1 i_2\cdots i_k i_{k+1}i_{k+2}\cdots i_N}
\epsilon^{j_1 j_2\cdots j_k j_{k+1}j_{k+2}\cdots j_N} 
Z_{i_1}^{j_1}Z_{i_2}^{j_2}\cdots Z_{i_k}^{j_k}
\ee
where $\cN_k^{-2}=k! (N-k)!$ is the normalization factor and $i,j$ indices run from one 
to $N$. According to BMN $\Tr Z^J$ corresponds to perturbative single string vacuum \cite{BMN} 
while $\cO_J$ is proposed to describe a single giant graviton state \cite{Bala-Strassler}. In 
the same spirit $n$-trace (or $n$-subdeterminant) operators of (\ref{chiral-primary-basis}a) 
(or (\ref{chiral-primary-basis}b)) corresponds to an $n$-strings (or $n$-giant gravitons) 
vacuum state.
Next we note that the last two operators in the row, 
$:(\Tr Z)^J:$, which according to BMN picture corresponds to $J$ string vacuum, and 
$:\cO_1^J:$, which corresponds to the $J$ tiny giant gravitons, are exactly equal
(note that $\cO_1=\Tr Z$).
This observation may help with a better understanding of the $X=0$ vacuum.}
We would like to emphasize that the $X=0$ vacuum 
classically does not look like a string theory vacuum state, but would become BMN vacuum 
quantum mechanically. In other words, we propose that the 't Hooft strings of our $U(J)$ gauge 
theory are indeed type IIB strings on the \pl\ \bg.
Our proposal for the $X=0$ vacuum has a counterpart in the BMN Matrix 
model where the $X=0$ (membrane) vacuum quantum mechanically behaves as a single five brane
vacuum \cite{MSV}. In section \ref{X=0Spectrum} we will provide another piece of evidence in 
support of the above proposal. 

In the BMN literature there have been two other quantum mechanical proposals for {\it 
perturbative} dynamics of fundamental type IIB closed strings. The first is obtained by 
summarizing the information coming from the ${\cal N}=4, D=4$ gauge theory calculations in the 
BMN sector, e.g. see \cite{Plefka-lectures, BMN-quantum-mechanices}. The 
second one is the 
Verlinde's String Bit model 
\cite{Stringbits}. We will comment on the possible relation between our $U(J)$ tiny graviton 
Matrix theory and the string bit model in section \ref{conclusion}.

\subsection{Spectrum of the theory about the vacua}\label{pertubative-spectrum}

In this section we start with studying the tiny graviton Matrix theory by performing a 
perturbative expansion about its vacua. Here we only consider $X=J$ and $X=0$ vacua and the 
analysis of rest of the vacua (which are in reducible \rep s of $Spin(4)$) are postponed to 
future works. The purpose of choosing these two vacua among the others is that, for $X=J$, 
which in string theory limit \eqref{string-theory-limit} should reproduce the single giant 
graviton state, the spectrum should match with the one given in \cite{Hedgehog}. This would 
provide us with another  check for our conjecture. In particular, there is a non-trivial, 
crucial test that our conjecture should pass: As we mentioned earlier, to obtain the \Ham , we 
started form the Born-Infeld action \eqref{DBI-action} in which we have not included the 
$U(1)$ gauge fields of the 3-brane. So, it is very important for the consistency of our theory 
to make sure that the photon states corresponding to this gauge field are present in the 
spectrum of our theory about $X=J$ vacuum. Although this sounds too good to be true,  performing 
explicit and detailed calculation we confirm the presence of these photon modes.

The reason to study the spectrum of the theory about the $X=0$ vacuum is to provide another 
piece of evidence in support of the conjecture we made in section 
\ref{vacuum-solutions-subsection}. As we proposed fundamental type IIB closed strings should
be realized as non-perturbative states about this vacuum. (This point would become clearer
in section \ref{effective-coupling} where we show that the effective coupling about this vacuum 
goes to 
infinity in the the string theory limit.) On the other hand, only the information about 
BPS states which are protected by supersymmetry can be trusted at strong coupling limit.
As we will show the spectrum of BPS states about this vacuum exactly matches with the spectrum 
of BPS type IIB closed string states in the \pl\ \bg .

\subsubsection{Irreducible vacuum}\label{X=JSpectrum}

In order to study the theory about the irreducible solution of (\ref{vacua-equations}a), we 
expand $X^i$'s as
\be\label{Xi-expansion}
X^i=\sqrt{\frac{\mu g_s}{R_-}} J^i +Y^i
\ee
where $J^i$'s are in irreducible $J\times J$ \rep s of $Spin(4)$ which satisfy
\be\label{equation-for-J}
[J^j,J^k, J^l,{\cal L}_5]=\epsilon^{ijkl} J^i\ ,
\ee
and $Y^i$'s parametrize perturbations about this vacuum.
Inserting \eqref{Xi-expansion} into \eqref{vacuum-potential} and keeping the terms second order 
in $Y^i$ and $X^a$ we have:
\be\label{V(2)-XJ}
V^{(2)}_{X=J}= \frac{\mu^2}{2R_-}\ \Tr \Biggl[
\left(Y^l+ \frac{1}{2}\epsilon^{ijkl} [ J^i , J^j , Y^k, {\cal L}_5]\right)^2 
+ X^2_a+ \frac{1}{2} [ J^i , J^j , X^a, {\cal L}_5][ J^i , J^j , X^a, {\cal L}_5]
\Biggr].
\ee

\vspace{2mm}
{\it Spectrum of $X^a$ modes}:
\vspace{2mm}

Using the by-parts integration property of the four brackets ({\it cf}. Appendix 
\ref{Quantum-Nambu}) the last term of \eqref{V(2)-XJ} can be written as
\[
\frac{1}{2}\Tr\left( X^a \bigl[ J^i , J^j , [ J^i , J^j , X^a, {\cal L}_5], {\cal 
L}_5\bigr]\right).
\]
Recalling the definition of $J^i$, namely \eqref{equation-for-J} and also \eqref{S3F-JJ}, 
it is straightforward to observe that the operator
\be\label{L^2}
\frac{1}{2}\bigl[ J^i , J^j , {\cal L}_5, [ J^i , J^j ,  {\cal L}_5], \star ]\bigr]
\ee
is in fact the second rank Casimir of $Spin(4)=SU(2)\times SU(2)$, and hence
\be\label{X^a_j}
\bigl[ J^i , J^j , {\cal L}_5, [ J^i , J^j ,  {\cal L}_5], X^a_j ]\bigr]=2j(j+2) X^a_j\ ,
\ee
where $X^a_j$ is the symmetric traceless rank $j$-tensor of $Spin(4)$, namely it is in 
$(\frac{j}{2}, \frac{j}{2})$ of $SU(2)\times SU(2)$ and $0\leq j\leq n$, where $n$ and $J$ are 
related as given in \eqref{size-dim}.
Using \eqref{X^a_j} one can read the eigen-frequencies of $X^a_j$, modes:
\be\label{Xa-mass}
\omega^2_j=\mu^2\left[1+\frac{1}{2}\cdot 2j(j+2)\right]=\mu^2 (j+1)^2\ \qquad\ 0\leq j\leq 
n .
\ee
Degeneracy of each of $X^a_j$ modes is $(j+1)^2$. In particular note that $j=0$ mode, with 
$\omega= \mu$, corresponds to center of mass motion of the giant three sphere.

\vspace{2mm}
{\it Spectrum of $Y^i$ modes and photon states}:
\vspace{2mm}

To diagonalize $Y^i$ part  of \eqref{V(2)-XJ}, consider the following eigenvalue equation:
\be\label{Y^i-modes}
\frac{1}{2}\epsilon^{ijkl}[ J^i , J^j , Y^k, {\cal L}_5]=\lambda Y^l\ .
\ee
Solutions to the above equation are eigenmodes of \eqref{V(2)-XJ} with eigen-frequencies 
$\omega^2=\mu^2(1+\lambda)^2$ or
\be\label{lambda-frequency}
\omega=\mu |1+\lambda|\ .
\ee

To solve \eqref{Y^i-modes}, we recall that $\frac{1}{2}\epsilon^{ijkl}[ J^i , J^j ,  {\cal 
L}_5, \star ]$ is indeed generator of $SO(4)$ rotations in $J\times J$ \rep . ({\it cf}. 
section 2 of \cite{Hedgehog}). Inspired by the continuum solutions, one can check that 
\eqref{Y^i-modes} has three class of solutions:

{\it i}) $\hat{Y}_j$-modes: for which $\lambda=j, \ \qquad 0\leq j\leq n$.
\newline
The ${\hat{Y}}_j$ modes are in $(\frac{j+1}{2}, \frac{j+1}{2})$ \rep\ of $SU(2)\times SU(2)$ 
and  have frequencies $\omega=\mu (j+1)$. Degeneracy of these modes is $(j+2)^2$.

{\it ii}) $\tilde{Y}_j$-modes: for which $\lambda=-(j+2), \ \qquad 1\leq j\leq n$.
\newline
The ${\tilde{Y}_i}$ modes are in $(\frac{j-1}{2}, \frac{j-1}{2})$ \rep\ of $SU(2)\times SU(2)$ 
with frequencies $\omega=\mu (j+1)$. ${\tilde{Y}_i}$ modes are $j^2$-fold degenerate.

{\it iii}) Zero modes, modes with $\lambda=-1$.
\newline
These modes are in 
$(\frac{j-1}{2}, \frac{j+1}{2})\oplus (\frac{j+1}{2}, \frac{j-1}{2})$ 
\rep\ of $SU(2)\times 
SU(2)$ with $1\leq j\leq n$ and have $\omega=0$.   These modes come with degeneracy $2j(j+2)$ 
and are gauge degrees of freedom (corresponding to the $SO(4)$ rotations of the fuzzy three 
sphere, {\it cf.}  section 2 of \cite{Hedgehog} and also section 5 of \cite{DSV1}). 

The physical modes, $\hat{Y}_j$ and $\tilde{Y}_j$ modes, both have the same frequency:
\be\label{Y-frequency}
\omega_j=\mu (j+1)\ .
\ee
Note, however that for $j=0$ we only have $\hat{Y}$ mode. These modes, which come with 
degeneracy equal to four, correspond to the center of mass motion of  the giant. The
$\tilde{Y}_{j=1}$ mode, however, corresponds to breathing mode of the giant, the mode in 
which  the giant maintains its spherical shape and only has radial fluctuations.

Beside the above eigenmodes, $Y^i$ modes also have two other physical eigenmodes solving the 
equation
\be\label{photon-eigen-equation}
\frac{1}{2}\epsilon^{ijkl}[ J^i , J^j , Y^k, {\cal L}_5]=\pm\lambda {\cal L}_5 Y^l\ .
\ee
The eigenmodes resulting from \eqref{photon-eigen-equation}, which will be denoted by 
$A^{\pm}_j$, 
are in 
$(\frac{j-1}{2}, \frac{j+1}{2})$ and  $(\frac{j+1}{2}, \frac{j-1}{2})$ 
\rep\ of $SU(2)\times 
SU(2)$ with $1\leq j\leq n$ (respectively corresponding to $+$ 
and $-$ sign in \eqref{photon-eigen-equation}). Both of these modes have frequencies
\be\label{photon-frequency} 
\omega=\mu (j+1), \qquad 1\leq j \leq n\ .
\ee

$A^{\pm}_j$ modes exactly correspond to the two different polarization of photon states. 
This is compatible with the results of the 
continuum limit \cite{Hedgehog}. So, our model passes another crucial test. 
Note that these photon states have the same 
$SU(2)\times SU(2)$ \rep\ as the zero modes.\footnote{ One should note that, for the fuzzy 
three sphere case, unlike the $S^2_F$, the algebra of functions generated from coordinates, 
$J^i$'s, is {\it not} covering the whole $J\times J$ matrices. In fact 
$Mat_C(J)$ contains them with multiplicity two \cite{sunjay2}. So, it is not strange to have 
zero modes and photons in the same $SU(2)\times SU(2)$ \rep s.} 
Moreover, the photon state starts from $j=1$, and hence, there is no photon mode with the same 
frequency as the center of mass modes, $\hat{Y}_{j=0}, X^a_{j=0}$.
We have summarized the results of this subsection in Table 1.

\begin{table}[ht]
\begin{center}
\begin{tabular}{|c||c|c|c|c|}
\hline
{\rm Mode} & $SU(2)\times SU(2)$ rep. & {\rm Frequency}  &   {\rm Degeneracy} &  {\rm Range}
\\ \hline\hline
$\hat{Y}_{j}$ & $(\frac{j+1}{2}, \frac{j+1}{2})$ & $\mu(j+1) $ &  $(j+2)^2$ & $0 \leq j \leq n$
\\ \hline
 $\tilde{Y}_{j}$ & $(\frac{j-1}{2}, \frac{j-1}{2})$ & $\mu(j+1) $ &  $j^2$ & $1 \leq j \leq n$
\\ \hline
 ${X}^a_{j}$ & $(\frac{j}{2}, \frac{j}{2})$ & $\mu(j+1) $ &  $4\cdot(j+1)^2$ & $0 \leq j \leq 
n$
\\ \hline
 $A^+_{j}$ & $(\frac{j-1}{2}, \frac{j+1}{2})$ & $\mu(j+1) $ &  $j(j+2)$ & $1 \leq j \leq n$
\\ \hline
 $A^-_{j}$ & $(\frac{j+1}{2}, \frac{j-1}{2})$ & $\mu(j+1) $ &  $j(j+2)$ & $1 \leq j \leq n$
\\ \hline
\end{tabular} \caption{Some of the modes and masses for the irreducible $X=J$ vacuum}
\end{center}\label{table1}
\end{table}
As we see we have $8\cdot (j+1)^2$ modes of frequency $\mu (j+1)$. It is straightforward to 
observe that all the modes in Table 1 are physical, in the sense that they all satisfy the 
Gauss law constraint \eqref{Gauss-law}.
These modes fall into the 
same \super \rep . In a similar manner, one may also work out spectrum of fermionic modes which 
we do not present it here. The detailed analysis of the spectrum and its superalgebra \rep s
are postponed to future works. (A similar analysis for the BMN Matrix model can be found in 
\cite{DSV2}.)

The number of the mode that we have listed in the Table 1 together with corresponding 
fermionic modes, is $\sum_{j=1}^n 16(j+1)^2$, which 
for large $n$ grows like $n^3$. The size of our matrices are, however, $J^2\sim n^4$. This 
means that besides the above  modes there are many other modes which we have not discussed 
here. Studying the structure of our $J\times J$ matrices and how they are filled up with 
$Spin(4)$ \rep s, suggests that (at least some of) these extra modes, which are in $(j_1, 
j_2),\  |j_1-j_2|\geq 2$ of $SU(2)\times SU(2)$, are  of the form 
of  the spike solutions of \cite{Hedgehog}, which in the continuum limit become the 
excitation modes of open strings ending on the giant and some of these mode would become 
massive and decouple in the contimuum limit. 
This of course deserves a more thorough study, postponed to future works.  

\subsubsection{$X=0$ vacuum}\label{X=0Spectrum}

In this case since the vacuum configuration is $X=0$, the $X^i$ and $X^a$'s can directly be 
regarded as perturbations about this vacuum. The \Ham\ \eqref{Matrix-model-Ham} up to second 
order in perturbations is
\be\label{H2-X=0}
{\bf H}^{(2)}_{X=0}= \Tr\Biggl[ 
\left(\frac{R_-}{2}\Pi_i^2+ \frac{\mu^2}{2R_-} X^2_i\right) +\left(\frac{R_-}{2}\Pi_a^2+ 
\frac{\mu^2}{2R_-} X^2_a\right)+
{\mu} \psi^\dagger {}^{\alpha \beta} \psi_{\alpha \beta}-
{\mu} \psi_{\dot\alpha \dot\beta}\psi^\dagger {}^{\dot\alpha \dot\beta} 
\Biggr]\ .
\ee
\eqref{H2-X=0} is of the form of $16=2\cdot 8$ decoupled harmonic oscillators, all with 
frequency 
$\mu$. Hence, one can readily read the spectrum. However, let us first introduce the creation 
operators
\begin{equation}
\label{creation-opts}
a^\dagger_i=\sqrt{\frac{\mu}{2R_-}} X^i- i \sqrt{\frac{R_-}{2\mu}} \Pi_i,\ \qquad
a^\dagger_a=\sqrt{\frac{\mu}{2R_-}} X^a- i \sqrt{\frac{R_-}{2\mu}} \Pi_a\ .
\end{equation}
The lowest energy states about this vacuum are gathered in Table 2.
\begin{table}[ht]
\begin{center}
\begin{tabular}{|c||c|c|}
\hline
{\rm Mode} &  {\rm Energy}  &   {\rm Degeneracy} 
\\ \hline\hline
$|0\rangle$ &    $0$  & 1 
\\ \hline
$\Tr a^\dagger_i |0\rangle$ & $\mu$ & 4  
\\ 
$\Tr a^\dagger_a |0\rangle$ & $\mu$ & 4  
\\
$\Tr \psi^{\dagger\alpha\beta} |0\rangle$ & $\mu$ & 4  \\
$\Tr \psi^{\dagger \dot\alpha\dot\beta} |0\rangle$ & $\mu$ & 4  
\\ \hline
\end{tabular} \caption{Lowest energy states about $X=0$ vacuum}
\end{center}
\end{table}
One may also construct two and higher oscillator states, e.g. 
$\Tr a^\dagger_i\Tr a^\dagger_j|0\rangle$ or $\left(\Tr (a^\dagger_i 
a^\dagger_j)-\frac{1}{J} \Tr a^\dagger_i\Tr a^\dagger_j \right)|0\rangle$ which have energy 
$2\mu$. 

The important point here is that: the action \eqref{H2-X=0} is basically the same as the \pl\ 
string theory $\sigma$-model action, once we ignore the string tension term (i.e. large $\mu$ 
limit). This in particular implies that the BPS states of strings on the \pl\ \bg , which 
necessarily have zero  worldsheet momentum, have exactly the same spectrum as the one generated 
by \eqref{H2-X=0}. In other words \eqref{H2-X=0} reproduces the spectrum of type IIb 
supergravity on the \pl\ \bg, this can be explicitly checked comparing spectrum of 
\eqref{H2-X=0} with TABLE I of \cite{review}.

\subsection{Effective coupling of the theory}\label{effective-coupling}   

In the previous section we studied the spectrum of quadratic parts of \Ham\ about $X=J$ and 
$X=0$ vacua, assuming that the cubic and higher order terms are small. To verify this 
assumption we need to keep the cubic and higher order terms in the expansion and obtain the 
coefficient in front of these terms, the coupling of the theory. However, first we should 
bring the quadratic parts of the \Ham\ to the ``canonical'' form
\[
H^{(2)}\sim \sum \omega_j {\cal A}^\dagger_j {\cal A}_j
\]
where ${\cal A}_j$'s are generic creation operators satisfying $[{\cal A}_j, 
{\cal A}^\dagger_{j'}]=\delta_{jj'}$
(note that this commutator is a quantum operator commutator and not a matrix commutator).
This can be done by rescaling $X$'s as
\be\label{rescale}
X^i\to \sqrt{\frac{R_-}{\mu}} X^i\ , \qquad X^a\to \sqrt{\frac{R_-}{\mu}} X^a,
\ee
(no scaling for fermions is needed). One can then observe that the coefficient in front of 
order $m$-terms ($m=3,\cdots, 6$)  is $\left(\frac{R_-}{\mu\sqrt{g_s}}\right)^{m-2}$. 
Therefore, the 
``bare'' coupling of the theory is:
\be\label{bare-coupling}
g^2_{bare}=\frac{R_-^2}{\mu^2 g_s}\ .
\ee
The reason we have called the above coupling, the bare coupling is that generically 
 there are dressing factors (power of $J$) which are multiplied to $g^2_{bare}$. The effective 
coupling and the dressing factors of course depend about which vacuum we are expanding the 
theory (for a similar discussion on BMN Matrix model see \cite{DSV1}). As we see from 
\eqref{bare-coupling}, we can trust our perturbation theory analysis when $g_{bare}\ll 1$. That 
is we are working in the large $\mu$ limit and performing a $1/\mu$ expansion.

\vspace{2mm}
{\it Effective coupling about  $X=0$ vacuum}:
\vspace{2mm}
 
Here we deal with a $U(J)$ gauge theory. Despite the fact our gauge theory is not a Yang-Mills 
theory, we expect the 't Hooft arguments about the planar and non-planar expansion to still 
hold because, 't Hooft's arguments are based on combinatorics of $J\times J$ matrices and not 
details of the model.  Therefore, we expect the effective coupling of the theory about this 
vacuum to be the 't Hooft coupling:
\be\label{X=0-coupling}
g_{X=0}=g^2_{bare} J =\frac{J^3}{(\mu p^+)^2 g_s}\ .
\ee 
The above should, however, be verified through explicit calculations with our model, e.g. by 
computing corrections to the spectrum of our states, employing time-independent quantum 
mechanical perturbation theory. Then, $g_{X=0}$, and not $g_{bare}$, should appear in the mass 
corrections.

In the string theory limit \eqref{string-theory-limit}, the effective coupling $g_{X=0}$ blows 
up ($\sim J^3$) and hence in this limit the $X=0$ vacuum is quite non-trivial and strongly 
coupled and one cannot trust perturbative information, expect for the BPS (protected) data.

\vspace{2mm}
{\it Effective coupling about  $X=J$ vacuum}:
\vspace{2mm}

Working out the dressing factors here are tricker than the previous case and we need to develop 
an extra knowledge about $Spin(4)$ \rep s and its Clebsch-Gordon coefficients. However, noting 
that $Spin(4)=SU(2)_L\times SU(2)_R$ paves the way to use the more standard information 
of $SU(2)$ Clebsch-Gordon coefficients and in particular Wigner $3j$ and $6j$ symbols (e.g. see 
\cite{6j-symbols}). A similar analysis has been done in \cite{DSV1}. For large $J$, and hence 
large $n$, where $n$ is the highest spin of either of the $SU(2)_L$ or $SU(2)_R$, (recall that 
$J\sim n^2$ \eqref{size-dim}), these $6j$-symbols lead to a factor of $n^{-3/2}$.  
However, one should remember that we have two $SU(2)$'s and there is a factor of 
$1/\sqrt{n}$ coming from the normalization factors of the other $SU(2)$. So, altogether we 
expect to have a dressing factor of $\frac{1}{n^2}\sim \frac{1}{J}$ and consequently
\be\label{X=J-coupling}
g_{X=J}=g_{bare} \frac{1}{J}=\frac{R_-}{\mu \sqrt{g_s}}=\frac{1}{\mu p^+\sqrt{g_s}}\ .
\ee
This is compatible with the results of \cite{Hedgehog}. It is interesting to note that
$g^2_{X=J}$ is indeed inverse of the effective coupling of strings on the \pl\ \bg $g_2$ (
in terms of BMN gauge theory parameters  $g_2=J^2/N$) \cite{Hedgehog}. In the string theory 
limit $g_{X=J}$ remains finite and hence the giant graviton theory is a perturbatively 
accessible theory.

\section{Conclusions and Discussion}\label{conclusion}

In this paper we conjectured that type IIB string theory on the \pl\ \bg , in the DLCQ 
description, is governed by a $0+1$ dimensional gauge theory, the tiny graviton Matrix theory,  
with the \Ham\  \eqref{Matrix-model-Ham}. In this theory, which is the \holo\ description of 
strings on the \pl ,  fundamental objects are the tiny 
gravitons, which are spherical 3-branes of very small size. We gave preliminary set of evidence 
in support of this conjecture,  analyzed vacua of the theory and discussed a continuum limit 
in which we recover the ten dimensional string theory.

There is another description of the same theory, the BMN gauge theory, in which,  unlike our 
case,  fundamental closed strings are perturbative.
Following the procedure of taking the Penrose limit and recalling the analysis of BMN 
more carefully we observe that in terms of ${\cal N}=4, D=4, \ U(N)$ SYM parameters
$R_-/\mu=\sqrt{g_s N}$ and hence the fuzziness parameter \eqref{fuzziness-l}, in the ten 
dimensional Planck units, is
\be\label{Rtiny-l}
l^2=\sqrt{\frac{1}{N}} \lp^2 =R^2_{tiny} \ .
\ee
For the second equality we have used \eqref{AdSR-planck-units} and \eqref{tiny-planck-AdS}.
This result is very interesting: the same parameter which controls the fuzziness, $l$, which is 
also the size of tiny gravitons, is exactly the same as the parameter  governing string 
interactions, $1/N$. The above together with \eqref{string-theory-limit} suggests that the 
smooth string picture in the \pl\ or 
$AdS_5\times S^5$ \bg\ only emerges in the $N\to\infty$ limit. 

Here we also briefly comment on the possible connection between our model about $X=0$ vacuum 
and the 
string bit model \cite{Stringbits}, and also a way which would hopefully help to reconstruct 
the tension term in the \lc\ 
string $\sigma$-model action. In string bit model strings are composed of $J$ bits, instead of 
$J\times J$ matrices. The string bit model is a supersymmetric quantum mechanics of $J$ 
particles with \super superalgebra built in \cite{Stringbits}.
A possible connection to our Matrix model may come  from identifying the string bit coordinate 
with the diagonal elements of our matrices. It is easy to see that a diagonal matrix of the 
form
\be\label{diagonal-matrix}
X^I=\left(\begin{matrix}
  x^I_1&       &    &   &   &  \cr
       & x^I_2 &    &   &   &  \cr
       &      &\cdot&   &   &  \cr
      &       &    &\cdot&  &  \cr
      &  &  & &\cdot &  \cr
     &  &  & & & x^I_J  \cr
\end{matrix}\right)\ ,
\ee
can be a classical solution (if $x^i$'s  are solutions to harmonic oscillator with frequency $\mu$). In 
addition, \eqref{diagonal-matrix} is not a gauge invariant 
solution of our Matrix model. (Under a generic $U(J)$ gauge 
tranformation it loses its diagonal form.) However, if we restrict our gauge transformations 
to ${\cal S}_J\subset U(J)$ (${\cal S}_J$ is the group of permutation of $J$ objects), the 
diagonal 
matrices remain diagonal. Interestingly, ${\cal S}_J$ is the symmetry group of the string bit 
model \cite{Stringbits}. Hence one may hope that it should be possible, to recover string bit 
model once we make the restriction to ${\cal S}_J$ gauge transformation, or to argue that 
quantum mechanically $U(J)$ symmetry is naturally reduces to ${\cal S}_J$.

Moreover, presumably the quadratic parts of the string bit model \Ham\ would then appear as the 
one loop effective action in our Matrix model, once we integrate out next to diagonal elements 
of matrices about the strong coupling fixed point. In other words, the string tension terms, 
which in the discrete format takes the 
form $\sum\frac{1}{J}(x^I_{j+1}-x^I_j)^2$, stems from the interactions between the tiny 
gravitons. (Note also that in our quantization  $J^{-1}$ play the role of $\hbar$ and as the 
tension term comes with a factor of $1/J$ it is reasonable to think of it 
as a term in the one loop effective action.)

It has been shown that
the \pl\ with a compact $X^-$ naturally appears as a certain Penrose limit of $AdS_5\times 
S^5/Z_k$ orbifold in the large $k$ limit \cite{Deconstruct}. As the 
$AdS_5\times S^5/Z_k$ is dual to an ${\cal N}=2$ $SU(N)^k$ super-conformal gauge theory,
this implies that DLCQ of strings on the \pl\ has a natural description in the ``BMN sector''
of this gauge theory. It would be nice to make connection between our formulation and the
BMN sector of the ${\cal N}=2$ $D=4$ super-conformal gauge theory .

The other interesting question is that whether our proposal is good enough to provide us with a 
$0+1$ dimensional gauge theory description for DLCQ of type IIB strings in the flat space. At 
the level of 
the \Ham\ \eqref{Matrix-model-Ham}, flat space limit, i.e. $\mu\to 0$ limit, seems to be a 
smooth one. However, as we have shown in section \ref{effective-coupling}, all the vacua of the 
Matrix model become strongly coupled and our perturbative $1/\mu$ expansion breaks down. 
Therefore, although this limit exists, the theory becomes intractable. It would also be 
interesting to understand how our Matrix model at $\mu=0$ and the other type IIB string Matrix 
theories, the IKKT \cite{IKKT} and Susskind-Sethi \cite{SS} model, are related to each other.
(The former is a quantum mechanics of instantons $0+0$ gauge theory and the latter is a 
$2+1$ SYM.) 

Finally we would like to comment on whether our method for quantization of 
of Nambu odd brackets can be used for quantizing a six dimensional $(0,2)$ theory, the theory 
which lives on spherical M5-brane giants. If our method is applicable, similarly to the 
3-brane case, the quantized M5-brane theory should become a theory on a fuzzy five sphere.
This idea, however, does not pass the first test: the relation between the radius of a $S^5_F$ 
and the size of matrices describing it is $\left(\frac{R}{l}\right)^5\sim J$ ({\it cf.} 
\eqref{radius-size}), whereas for a five sphere giant $\left(\frac{R}{l}\right)^4\sim \mu p^+$
\cite{MST, MSV}. Of course this argument does not completely rule out the
applicability of this idea and this direction deserves further analysis.

{\large{\bf Acknowledgements}}
\newline
I would like to thank Mohsen Alishahiha, Keshav Dasgupta, Jan Plefka, Darius Sadri, Lenny 
Susskind, Steve 
Shenker and Scott Thomas for  comments.
The work of M. M. Sh-J. is supported in part by NSF grant
PHY-9870115 and in part by funds from the Stanford Institute for Theoretical
Physics. 

 
\appendix
                                                                                                    
\section{Light-cone Hamiltonian of a 3-brane in the plane-wave background}\label{LCHam}

As we discussed our tiny graviton matrix theory conjecture is based on the light-cone 
quantization (regularization) of a 3-brane action in the maximally supersymmetric plane-wave 
background.  
In this Appendix we review, essentially quoting from \cite{Hedgehog}, how to obtain the \lc\ 
\Ham . We start with the Born-Infeld action for a 3-brane:
\be \label{DBI-action}
  S=  \frac{1}{l_s^4g_s}\int d\tau d^3\sigma \:
  \sqrt{-\det \left( G_{\mu\nu}\partial_{\hmu} X^{\mu}\partial_{\hnu} X^{\nu} \right)} +\int C_4
\ee
where $\hmu,\hnu=0,1,2,3$ indices correspond to the  worldvolume coordinates $\tau,
\sigma^r$, $r=1,2,3$. $X^\mu$ are the embedding coordinates and $G_{\mu\nu}$ and $C_4$ are the
background plane-wave metric and fourforms \eqref{background} and \eqref{fourform}. 
Note that we have {\it not} included the $U(1)$ gauge field of the brane. In the whole paper we 
set $l_s=1$ and unless it is emphasized explicitly, all our quantities are measured in units of 
$l_s$.

To fix the light-cone gauge we set
\begin{subequations}\label{light-cone}
\begin{align}
X^+&=\tau\\
g_{0r}&\equiv G_{\mu\nu} \frac{\partial X^\mu}{\partial \tau}\frac{\partial X^\nu}{\partial 
\sigma^r}=0
\end{align}
\end{subequations}
The momentum conjugate to $X^+,\ 
\frac{\partial {\cal L}}{{\partial \partial_\tau X^+}}$ , is the light-cone \Ham\  $H_{lc}$
and  the momentum conjugate to $X^-,\ 
\frac{\partial {\cal L}}{{\partial \partial_\tau X^-}}$ , is the light-cone momentum  $p^+$. 
The momentum conjugate to the directions transverse to the light-cone, $X^I,\ I=1,2,\cdots , 8$ 
is $P_I=\frac{\partial {\cal L}}{{\partial \partial_\tau X^I}}$.

Fixing the light-cone gauge fixes a part of the diffeomorphism invariance of the Born-Infeld 
action which mixes the worldvolume time, $\tau$, with the worldvolume spatial directions 
$\sigma^r$.  The part of the diffeomorphisms which only act on the spatial directions, however, 
are still present and not fixed. It is noteworthy that as always in the \lc\ gauge fixing there 
is no need of introducing ghosts (the ghosts are decoupled).

Simplifying the \lc\ \Ham, we obtain \cite{Hedgehog}
\be\label{Hlc-bosonic}
\begin{split}
{H}^{bos.}_{l.c.} = \int d^3\sigma& \Biggl[\frac{1}{2 p^+}(P_i^2+ P_a^2)+\frac{\mu^2 
p^+}{2}(X_i^2+X_a^2)+\frac{1}{2 p^+g_s^2} \; \mbox{det}g_{rs}\cr
&- \frac{\mu}{6g_s} \Bigl(
              \epsilon^{i j k l} X^i\{ X^j, X^k, X^l \}+ \epsilon^{a b c d} X^a\{ X^b, X^c,
X^d \} \Bigr)\Biggr] \ ,
\end{split}
\ee
where
\[
g_{rs}=\partial_r X^i\partial_s X^i+\partial_r X^a\partial_s X^a\ ,
\] 
and the brackets are Nambu three brackets \cite{Nambu} and are defined as
\be\label{Nambubracket}
         \{ F , G , K \} = \epsilon^{p r s} \partial_p F \partial_r G \partial_s K  ,
\ee
where $F, G$ and $K$ are arbitrary functions of the worldvolume coordinates $\sigma^r$. The
$\det g_{rs}$ term can also be written in terms of Nambu brackets \cite{Squashed}
\bea\label{det}
     \mbox{det}g_{rs} &=&  \frac{1}{3!} \{ X^i , X^j , X^k \} \{ X^i , X^j , X^k \}
+\frac{1}{3!} \{ X^a , X^b , X^c \} \{ X^a , X^b , X^c \} \cr &+&
\frac{1}{2!} \{ X^i , X^j , X^a \} \{ X^i , X^j , X^a \}
+\frac{1}{2!} \{ X^i , X^a , X^b \} \{ X^i , X^a , X^b \}.
\eea
The first line of \eqref{Hlc-bosonic} is coming from the Born-Infeld term and the second line 
from the Chern-Simons term ($C_4$ term) of \eqref{DBI-action}. The term proportional to $\mu^2$ 
comes from the $(dx^+)^2$ term of the metric \eqref{background} and the terms linear in $\mu$ 
from the fiveform of the background.

The above action, besides the diffeomorphism on $\sigma^r$ directions (which is a local gauge 
symmetry), has an $SO(4)\times SO(4)$ global symmetry, the $SO(4)$'s acting on $X^i$ or $X^a$ 
directions, respectively. Moreover, there is a $Z_2$ symmetry which exchanges $X^i$ and $X^a$ 
directions \cite{Hedgehog}.
The condition (\ref{light-cone}b),
\be\label{level-matching}
p^+\partial_r X^-\approx P_i \partial_r X^i+ P_a \partial_r X^a 
\ee
where $\approx$ is the ``weak'' equality, should be imposed as a constraint on the dynamics of 
the system (or on the states, in the quantized version). Hence the dynamics of $X^-$ in the \lc\ 
gauge is completely determined through dynamics of the transverse directions.

The Hamiltonian \eqref{Hlc-bosonic} is the bosonic part of the full supersymmetric \Ham\ with
superalgebra \super \cite{Hedgehog, review}. The fermionic part of the \Ham\ after fixing the 
$\kappa$ symmetry in the light-cone gauge is \cite{Metsaev-D3, Hedgehog} 
\be\label{Hlc-fermionic}
\begin{split}
H_{l.c.}^{fer}
= \int d^3\sigma \Biggl[&\mu\ \psi^\dagger {}^{\alpha \beta} \psi_{\alpha \beta}+
\frac{2}{p^+ g_s}\left( 
  \psi^\dagger {}^{\alpha \beta} (\sigma^{ij})_\alpha^{\: \: \: \delta} 
  \{ X^i, X^j, \psi_{\delta \beta} \} +
  \psi^\dagger {}^{\alpha \beta} (\sigma^{ab})_\alpha^{\: \: \: \delta} 
  \{ X^a, X^b, \psi_{\delta \beta} \}\right)\cr
-&\mu\ \psi_{\dot\alpha \dot\beta}\psi^\dagger {}^{\dot\alpha \dot\beta} 
-\frac{2}{p^+ g_s}\left( \psi_{\dot\delta \dot\beta} 
(\sigma^{ij})_{\dot\alpha}^{\: \: \: \dot\delta} 
  \{ X^i, X^j, \psi^\dagger {}^{\dot\alpha \dot\beta}\}+
\psi_{\dot\delta \dot\beta} (\sigma^{ab})_{\dot\alpha}^{\: \: \: \dot\delta} 
  \{ X^a, X^b, \psi^\dagger {}^{\dot\alpha \dot\beta}\}\right)
 \Biggr]  .
\end{split}
\ee
Following the notation of \cite{review}, the fermions $\psi$ are
spinors of two different $SU(2)'s$, one coming from the decomposition of each of the
two $SO(4)$'s into $SU(2) \times SU(2)$; in other words, $\psi$ carries
two spinor indices, each being the Weyl index of one of the $SO(4)$'s. 
The fact that in type IIB both of fermions have the same ten dimensional chirality
 is reflected in the fact that both of our fermions should have 
the same $SO(4)$ chiralities \cite{review}.
The explicit mass terms (terms proportional to $\mu$) in $H^{fer}_{l.c.}$ for the fermions come 
from the contribution of RR fiveforms to the super-vielbeins.

\section{Nambu brackets and their quantization}\label{Quantum-Nambu}

As we showed in  Appendix \ref{LCHam}, Nambu 3-brackets naturally appear in the \lc\ \Ham\ of a
3-brane. In fact, following the logic of previous Appendix, it is straightforward to 
show that in general a Nambu $p$-bracket appears in the $p$-brane \lc\ \Ham ; for the case of M2 
and M5 -branes  this has been explicitly shown \cite{DSV1, MSV, Hoppe}.      
In this Appendix, first we define a general Nambu $p$-bracket and list its properties and then 
discuss quantization of brackets.

\subsection{Properties of Nambu brackets}

In general a Nambu $p$-bracket \cite{Nambu} is defined among $p$ functions $F_i(\sigma^r),\ 
i=1,2,\cdots, p$, where $F_i$ are (real) functions on a $p$-dimensional space, parametrized with
$\sigma^r,\ r=1,2,\cdots, p$:
\be\label{Nambu-p-bracket}
\{ F_1, F_2,\cdots, F_p\}\equiv 
\epsilon^{r_1r_2\cdots r_p} 
\frac{\partial F_1}{\partial \sigma^{r_1}} 
\frac{\partial F_2}{\partial \sigma^{r_2}} 
\cdots
\frac{\partial F_p}{\partial \sigma^{r_p}} . 
\ee
For $p=2$ the above reduces to the usual Poisson bracket.
The above bracket has five important properties: 

{\it i)} Cyclicity
\be\label{exchange}
\{ F_1, F_2,\cdots, F_i,\cdots, F_j,\cdots, F_p\}=-
\{ F_1, F_2,\cdots, F_j,\cdots, F_i,\cdots, F_p\}.
\ee
Consequently, under the cyclic rotation of functions inside bracket:
\be\label{cyclicity}
\{ F_1, F_2,\cdots, F_p\}=(-1)^{p+1}
\{ F_2, F_3,\cdots, F_p, F_1\}.
\ee
Eq.\eqref{cyclicity} is a consequence of a similar property of the $\epsilon$ symbol, 
namely
\be\label{cyclic-epsilon}
\epsilon^{r_1r_2\cdots r_p}=(-1)^{p+1}\epsilon^{r_2\cdots r_p r_1}. 
\ee
Therefore, even and odd Nambu brackets have a different cyclicity behaviour. As we will see in 
the next subsection this difference is the source of difficulties with quantization of Nambu odd 
brackets.

{\it ii)} Jacobi Identity
\be\label{Jacobi-classic}
\epsilon^{i_1 i_2\cdots i_{2p-1}} 
\bigg{\{} F_1, F_2,\cdots, F_{p-1},\{ F_p, F_{p+1},\cdots, F_{2p-1}\}\bigg{\}}=0.
\ee

{\it iii)} Associativity
\be\label{Associativity}
\{ F_1, F_2,\cdots, F_{p-1}, F_p G_p\}=\{ F_1, F_2,\cdots, F_{p-1}, F_p\} G_p+
F_p\{ F_1, F_2,\cdots, F_{p-1}, G_p\} .
\ee

{\it iv)} Trace property
\be\label{zero-trace}
\int d^p\sigma \{ F_1, F_2,\cdots, F_{p-1}, F_p \}=0
\ee

{\it v)} By part integration
\be\label{nambu-bypart}
\int d^p\sigma \{ F_1, F_2,\cdots, F_{p-1}, F_p \} G_p=
-\int d^p\sigma \{ F_1, F_2,\cdots, F_{p-1}, G_p \} F_p.
\ee
All the above properties can be verified directly from the definition \eqref{Nambu-p-bracket}. 
Note also that the integration by part property is a result of ${\it iii)}$ and ${\it iv)}$.

\subsection{Quantization of Nambu brackets}

The formulation of standard quantum mechanics is obtained from the usual Hamilton-Jacobi 
formulation of classical mechanics once we replace the Poisson brackets (i.e. Nambu two 
brackets)
with their ``quantized'' version, the matrix (or operator) commutators:
\be\label{commutator}
\{ A, B\} \longleftrightarrow \frac{1}{i\hbar} [\hat{A}, \hat{B}].
\ee
Since Nambu introduced his brackets \cite{Nambu} their quantization has been extensively studied 
and in particular quantization of Nambu odd brackets has appeared to be very challenging e.g. 
see \cite{Zachos}.

\subsubsection{Quantization of Nambu even brackets}\label{even}

{\it Prescription:}

\ \ \ \ {\it i)} Replace functions $F_i(\sigma)$ with operators (matrices) $\hat{F_i}$.

\ \ \ \ {\it ii)} 
\be\label{QN2p}
\{ F_1, F_2,\cdots, F_{2p}\} \longleftrightarrow 
\frac{1}{{i}^p} [
\hat{F_1},\hat{F_2},\cdots, {\hat{F}}_{2p}]\equiv 
 \frac{1}{i^p(2p)!}
\epsilon^{i_1i_2\cdots i_{2p}}
\hat{F_{i_1}}\hat{F_{i_2}}\cdots {\hat{F}_{i_{2p}}} 
\ee

\ \ \ \ {\it iii)}  
\be\label{int-trace}
\int d^{2p} \sigma\  * \longleftrightarrow Tr *
\ee
As wee see from \eqref{QN2p} the $2p$-bracket can be written in terms of a sum of products of 
$p$ commutators. Note also that, similarly to $\hbar$ in \eqref{commutator}, generically we need 
to introduce a ``quantization'' parameter in \eqref{QN2p}.

{}Using definition \eqref{QN2p} one can easily check that the Jacobi identity 
\eqref{Jacobi-classic} is satisfied, however, for $p>1$ the associativity is compromised
\cite{Zachos}. The trace property \eqref{zero-trace}, follows from \eqref{cyclic-epsilon}, the 
cyclicity of the trace and eq.\eqref{int-trace}.  
The by part integration property, despite the absence of associativity, is still satisfied.

\subsubsection{Quantization of Nambu odd brackets}\label{odd}

Noting \eqref{cyclic-epsilon}, it is easy to observe that if we use prescription \eqref{QN2p} 
for 
odd brackets we lose the trace and by part integration properties. These two properties are 
physically very important because they are connected with the (existence of) conserved charges.
Therefore, \eqref{QN2p} cannot be directly used for the odd brackets.

Here we propose a solution for the problem of quantization of Nambu odd brackets, which as we 
will see in the main text and also in Appendix \ref{S3F}, is 
a suitable one for the cases of our concern, namely  the cases where the $p$ brackets  come form 
the worldvolume of the $p$ dimensional branes.
The idea is to replace $2p-1$ brackets with  $2p$ brackets and again use \eqref{QN2p}. 
Explicitly:
\be\label{QN2p-1}
\{ F_1, F_2,\cdots, F_{2p-1}\} \longleftrightarrow 
\frac{1}{i^p} [
\hat{F}_1,\hat{F}_2,\cdots, {\hat{F}}_{2p-1}, \hat{{\cal L}}_{2p+1}]\equiv 
\frac{1}{i^p(2p)!} \epsilon^{i_1i_2\cdots i_{2p}}
{\hat{F}}_{i_1}{\hat{F}}_{i_2}\cdots {\hat{F}}_{i_{2p-1}} \hat{{\cal L}}_{2p+1} 
\ee
where $\hat{{\cal L}}_{2p+1}$ is a given fixed matrix. This matrix is closely 
related to the ``chirality'' operator for $2p$ dimensional fermions. In the case of our 
interest, namely, Nambu 
3-brackets where the explicit expression for \eqref{QN2p-1} is
\be\label{Nambu-3-bracket}
\begin{split}
[A,B,C,{\cal L}_{5}] &=\frac{1}{24}\Biggl(
[A,B][C,{\cal L}_{5}]-[A,C][B,{\cal L}_{5}]+[A,{\cal L}_{5}][B,C]\cr
&+[C,{\cal L}_{5}][A,B]-[B,{\cal L}_{5}][A,C]+[B,C][A,{\cal L}_{5}]\Biggr)\ ,
\end{split}
\ee
and ${\cal L}_{5}$ is made out of direct product of ${\bf 1}_{4\times 4}$ and $\Gamma^5$, 
where $\Gamma^5$ is the $SO(4)$ chirality operator. This point will be discussed in more 
detail in Appendix \ref{S3F}.

\section{Superalgebra of the plane-wave background}\label{SUSY-algebra}
 
The plane-wave solution \eqref{background} has a large set of bosonic and fermionic isometries. 
The bosonic isometry group, whose dimension is  30, contains  $SO(4)\times SO(4)$ and 
translation along $x^-$ and $x^+$ directions, the generators of which will be denoted
by ${\bf J_{ij}}$, ${\bf J_{ab}}$, $P^+$ and ${\bf H}$, respectively \cite{review}.

There are also 32 fermionic isometries (supercharges) which can be decomposed into 16 
kinematical 
supercharges 
$q_{\alpha\beta}$, $q_{\dot\alpha\dot\beta}$ (and their complex conjugates) and 16 dynamical 
supercharges $Q_{\alpha\dot\beta}$ and $Q_{\dot\alpha\beta}$. $P^+=i\frac{\partial}{\partial 
x^-}$
is at the center of the whole superalgebra, i.e. it commutes with all the bosonic and fermionic 
generators. The dynamical part of the superalgebra is (for the full superalgebra see 
\cite{review}):
\be\label{kinematical-susy}
\begin{split}
\{q_{\alpha\beta}, q^{\dagger\rho\lambda}\}  =\delta_\alpha^\rho\
\delta_{\beta}^{\lambda}\ P^+\ , &\quad
\{q_{\dot\alpha\dot\beta}, q^{\dagger\dot\rho\dot\lambda}\}  =\delta_{\dot\alpha}^{\dot\rho}\
\delta_{\dot\beta}^{\dot\lambda}\ P^+ ,\cr
[P^+, q]=0\ , &\quad [P^+, q^\dagger]=0\cr
[{\bf H}, q_{\alpha\beta}]=\mu q_{\alpha\beta}\ , &\quad [{\bf H}, 
q_{\dot\alpha\dot\beta}]=-\mu q_{\dot\alpha\dot\beta}\ .
\end{split}
\ee
\be\label{QPH}
\begin{split}
[P^+, Q]=0\ , &\quad [P^+, Q^\dagger]=0\cr
[{\bf H}, Q]=0\ , &\quad [{\bf H}, Q^\dagger]=0\ .
\end{split}
\ee
\be\label{susy-algebra}
\begin{split}
\{Q_{\alpha\dot\beta}, Q^{\dagger\rho\dot\lambda}\} & =\delta_\alpha^\rho\
\delta_{\dot\beta}^{\dot\lambda}\ {\bf {H}}
+2\mu (i\sigma^{ij})_{\alpha}^{\rho}
\delta_{\dot\beta}^{\dot\lambda}\ {{{\bf J}_{ij}}}
+2\mu (i\sigma^{ab})_{\dot\beta}^{\dot\lambda} \delta_{\alpha}^{\rho}\ {{{\bf J}_{ab}}}\cr
\{Q_{\dot\alpha\beta}, Q^{\dagger\dot\rho\lambda}\}& =\delta_{\dot\alpha}^{\dot\rho}\
\delta_{\beta}^{\lambda}\ {\bf {H}}
+2\mu (i\sigma^{ij})_{\dot\alpha}^{\dot\rho}
\delta_{\beta}^{\lambda}\ {{{\bf J}_{ij}}}
+2\mu (i\sigma^{ab})_{\beta}^{\lambda} \delta_{\dot\alpha}^{\dot\rho}\ {{{\bf J}_{ab}}}.
\end{split}
\ee
The above superalgebra \eqref{QPH} and \eqref{susy-algebra} (involving $Q$'s, ${\bf H}$ and 
$P^+$)  can 
be identified as 
$PSU(2|2)\times PSU(2|2)\times U(1)_- \times U(1)_+$ where the $U(1)_\pm$ are
translation along the $x^\pm$ directions. The bosonic part of $PSU(2|2)$ algebra
is $SU(2)\times SU(2)$ where each of $SU(2)$'s is coming from a different $SO(4)$ isometry of 
the background.

\subsection{ SUSY algebra in terms of $J\times J$ matrices}

The generators of the above supersymmetry algebra can be realized in terms of $J\times J$ 
matrices as
\begin{subequations}
\begin{align}
P^+&=\frac{1}{R_-} \Tr {\bf 1}\ ,\\
q_{\alpha\beta}=\frac{1}{\sqrt{R_-}}\ \Tr \psi_{\alpha\beta}\ \qquad ,& \qquad  
q_{\dot\alpha\dot\beta}=\frac{1}{\sqrt{R_-}}\ \Tr \psi_{\dot\alpha\dot\beta}\ .
\end{align}
\end{subequations}
\begin{subequations}
\begin{align}
{\bf J}_{ij}&= \Tr \left(X^i\Pi^j-X^j\Pi^i +
\psi^{\dagger\alpha\beta} (i\sigma^{ij})_{\alpha}^{\ \rho} \psi_{\rho\beta}
-\psi^{\dagger \dot\alpha\dot\beta} (i\sigma^{ij})_{\dot\alpha}^{\ \dot\rho} 
\psi_{\dot\rho\dot\beta}\right)\\
{\bf J}_{ab}&= \Tr \left(X^a\Pi^b-X^b\Pi^a +
\psi^{\dagger \alpha\beta} (i\sigma^{ab})_{\beta}^{\ \rho} \psi_{\alpha\rho}
-\psi^{\dagger \dot\alpha\dot\beta} (i\sigma^{ab})_{\dot\beta}^{\ \dot\rho}
\psi_{\dot\alpha\dot\rho}\right)
\end{align}
\end{subequations}
\be\label{Qsupercharge}
\begin{split}
Q_{\dot\alpha\beta}=\sqrt{\frac{R_-}{2}}\ \Tr &\Bigl[ 
(\Pi^i-i\frac{\mu}{R_-} X^i) (\sigma^{i})_{\dot\alpha}^{\ \rho} \psi_{\rho\beta}+
(\Pi^a-i\frac{\mu}{R_-} X^a) (\sigma^{a})_{\beta}^{\ \dot\rho} \psi_{\dot\alpha\dot\rho}\cr
+&
\frac{1}{3! g_s} \left(\epsilon^{ijkl}[ X^i, X^j, X^k, {\cal L}_5] 
(\sigma^{l})_{\dot\alpha}^{\ \rho}\psi_{\rho\beta}+
\epsilon^{abcd}[ X^a, X^b, X^c, {\cal L}_5] 
(\sigma^{d})_{\beta}^{\ \dot\rho}\psi_{\dot\alpha\dot\rho}\right)\cr 
+&
\frac{1}{2 g_s}\left( [ X^i, X^a, X^b, {\cal L}_5] (\sigma^{i})_{\dot\alpha}^{\ \rho}
(i\sigma^{ab})_{\beta}^{\ \gamma}\psi_{\rho\gamma}+
[ X^i, X^j, X^a, {\cal L}_5] (i\sigma^{ij})_{\dot\alpha}^{\ \dot\rho}
(\sigma^{a})_{\beta}^{\ \dot\gamma}\psi_{\dot\rho\dot\gamma}\right)\Bigr]
\end{split}
\ee
and similarly for $Q_{\alpha\dot\beta}$. The expression for ${\bf H}$  is given in 
\eqref{Matrix-model-Ham}.

The (anti)commutation relations \eqref{kinematical-susy}, \eqref{QPH} and 
\eqref{susy-algebra} may be verified 
using the quantum (as opposed to matrix) commutation relations:
\begin{subequations}
\begin{align}
[X^I_{pq}, \Pi^J_{rs}]=i\delta^{IJ}\ \delta_{ps}\delta_{qr}\ ,&\qquad I, J=1,2,\cdots, 8\cr
\{(\psi^{\dagger \alpha\beta})_{pq}, (\psi_{\rho\gamma})_{rs}\}=
\delta^{\alpha}_{\rho} \delta^{\beta}_{\gamma}\  
\delta_{ps}\delta_{qr}\, , & \ 
\{(\psi^{\dagger \dot\alpha\dot\beta})_{pq}, (\psi_{\rho\gamma})_{rs}\}=
\delta^{\dot\alpha}_{\dot\rho} \delta^{\dot\beta}_{\dot\gamma}\  
\delta_{ps}\delta_{qr}\ ,
\end{align}
\end{subequations}
where $p,q, r, s=1,2,\cdots , J$ are matrix indices.

As for the kinematical supersymmetry generators, here we have only presented $q$'s. The 
bosonic kinematical isometries can be worked out in a similar manner, through anti-commutation 
of
a dynamical supercharge and a kinematical one \cite{review}. The kinematical super-generators, 
are all in the $U(1)$ (trace) part of the $U(J)$ matrices. 

\section{A brief review on fuzzy $d$ spheres, $S^d_F$}\label{S3F}

Fuzzy spheres are (noncommutative) discretized spheres and the discretization (`` 
quantization'') of the sphere is performed in such a way that the $SO(d+1)$ rotation symmetry of 
a $d$ sphere, $S^d$, is preserved.
In order to give an idea how this  quantization can be done, recall that a round $d$ 
sphere of radius $R$ can be embedded in a $d+1$ dimensional flat space as
\be\label{radius}
\sum_{i=1}^{d+1} (X^i)^2=R^2.
\ee
Next  note that for any $d$ sphere the coordinates satisfy
\be\label{Nambu-sphere}
\{ X^{i_1}, X^{i_2},\cdots, X^{i_{d}}\}= R^{d-1} 
\epsilon^{i_1i_2\cdots i_{d+1}} X^{i_{d+1}} ,
\ee
where the left-hand-side is a Nambu $d$ bracket.

To see this more clearly let us consider the simple example of a two sphere, in which case the 
above bracket is essentially a Poisson bracket. Take the following embedding:
\[
X^1=R\sin\theta\cos\phi\ , \ \  
X^2=R\sin\theta\sin\phi\ , \ \  
X^3=R\cos\theta\ , 
\]
where $\theta,\ \phi$ are the spherical coordinates. It is then straightforward to check that
\[
\{X^1, X^2\}=\frac{1}{\sin\theta}\left(
\partial_\theta X^1\partial_\phi X^2-\partial_\theta X^2\partial_\phi X^1\right)=R^2\cos\theta=R 
X^3 ,
\] 
(Note that $1/\sin\theta$ factor is the volume-form on the $S^2$)  
and in general $\{X^i, X^j\}=R\epsilon^{ijk} X^k$. This point for $d=3, 5$ has been noted and 
used in \cite{Hedgehog, MSV}.
The advantage of using \eqref{radius} and \eqref{Nambu-sphere} is that they are both 
invariant under the $SO(d+1)$ rotations.

\subsection{ Fuzzifying an $S^d$} 

To quantize or ``fuzzify'' the $d$ sphere, we quantize the Nambu bracket \eqref{Nambu-sphere} 
using the prescription of quantizing  Nambu brackets discussed in Appendix \ref{Quantum-Nambu}.
The equation \eqref{radius} is then simply a condition on the size of the matrices. In other 
words, we replace $X^i$'s with the matrices which are finite dimensional \rep s of $SO(d+1)$ and 
\eqref{radius}, or the radius, is the second rank Casimir for that  \rep .  
In the process of quantization of Nambu brackets we need to introduce a parameter, $l$, which 
measures the amount of fuzziness. This parameter plays the role of $\hbar$ in the usual quantum 
mechanics and is defined in such a way that in the $l\to 0$ limit (while the radius is kept 
fixed) we recover the commutative round sphere.
To show how the quantization works and as a warm up we consider the case of fuzzy two sphere 
$S^2_F$. In this case the quantum version of \eqref{Nambu-sphere} is
\[
[X^i, X^j]=il\epsilon^{ijk} X^k
\]
where $l$ is the fuzziness and $X^i/l$ are generators of $SO(3)\simeq SU(2)$. 
The $S^2_F$ of radius $R$ is then given in terms of  a $J\times J$ \rep\  of $SU(2)$, where 
(e.g. see \cite{DSV1})
\be\label{fuzzytwosphere}
\left(\frac{R}{l}\right)^2=\frac{J^2-1}{4}\ .
\ee 
In order to fuzzify an even sphere we can simply generalize the above, using the arguments of 
Appendix \ref{Quantum-Nambu}. The case of fuzzy odd spheres, in which we face 
odd Nambu brackets, is trickier. 
A nice and detailed discussion on fuzzy spheres, even and odd,  may be found in 
\cite{sunjay1, sunjay2}. (We would like to note that the above method for quantizing or 
fuzzifying a sphere, which is quite generic and works for even and odd cases alike, 
is a new construction and to the author's knowledge it has not been presented in the literature 
previously. This method is built in such a way that in the continuum limit, $J\to \infty,\ 
R=fixed$, we recover the round commutative sphere.) 

Before moving to a more detailed discussion on the $S^3_F$ case, we  quote an important 
result about the fuzzy spheres. In general for a $S^d_F\ (d\geq 2)$ the size of the matrices, 
$J\times J$, and the radius $R$, for large $J$, are related as \cite{sunjay2}
\begin{subequations}\label{radius-size}
\begin{align}
\left(\frac{R}{l}\right)^{\frac{1}{8}d(d+2)} &\sim J \quad\quad\quad {\rm for\ even\ } d ,\\
\left(\frac{R}{l}\right)^{\frac{1}{8}(d-1)(d+5)} &\sim J \quad\quad\quad {\rm for\ odd\ } d .
\end{align}
\end{subequations}
For $d=2$ $\left(\frac{R}{l}\right)\sim J$ (reducing to \eqref{fuzzytwosphere}), for $d=3$
$\left(\frac{R}{l}\right)^2\sim J$  and  $\left(\frac{R}{l}\right)^3\sim J$ for $d=4$.

\subsection{Specific case of fuzzy three sphere $S^3_F$}
 
Now we focus on the case relevant to our tiny graviton Matrix model, the fuzzy three sphere, 
$S^3_F$. The $S^3_F$ is described through the embedding matrices $X^i$ which satisfy\footnote
{Another construction for $S^3_F/Z_2$ has been presented in \cite{Seif}.}
\begin{subequations}\label{S3F-def}
\begin{align}
[X^i, X^j, X^k, {\cal L}_5]&=-l^2 \epsilon^{ijkl} X^l\ , \\  
\sum_{i=1}^4 (X^i)^2 &= R^2\ .
\end{align}
\end{subequations}
To give an idea how to solve (\ref{S3F-def}a), let us start with the $4\times 4$ matrices, i.e.
\be\label{gamma-solution}
X^i=l\Gamma^i\ , \quad\qquad  {\cal L}_5=\Gamma^5,
\ee 
where $\Gamma$'s are the standard four dimensional Dirac matrices. It is evident that
\eqref{gamma-solution} solves (\ref{S3F-def}) for $R=2l$. In other words
\eqref{gamma-solution} defines a $S^3_F$ of radius two.
The necessity of the existence of ${\cal L}_5$ in the definition of \eqref{QN2p-1} or 
(\ref{S3F-def}) is clearly visible, if we demand to have a solution in the form of $\Gamma$ 
matrices.\footnote{Had we taken $X^i=l\Gamma^i \Gamma^5$, 
\[[X^i, X^j, X^k, {\cal L}_5]=+l^2 \epsilon^{ijkl} X^l. \]
Based on the discussions of  section \ref{Evidence}, this choice corresponds to an anti fuzzy 
sphere, 
i.e. a $S^3_F$ with orientation opposite to that of $X^i=l\Gamma^i$ solution. Anti-giants are 
solutions of \Ham\ \eqref{Matrix-model-Ham} with $\mu\to -\mu$.}

To construct a fuzzy three sphere of generic radius, we start with 
matrices made out of a sequence of $n$ $\Gamma$'s and $4\times 4$ identity 
matrices \cite{sunjay1,sunjay2} 
\begin{subequations}\label{LL5}:
\begin{align}
L^i&\equiv(
\Gamma^i \otimes {\bf 1}   \otimes {\bf 1}\otimes \cdots \otimes {\bf 1}+ {\bf 1}\otimes 
\Gamma^i 
\otimes \cdots \otimes {\bf 1} + \cdots + {\bf 1}\otimes {\bf 1}\otimes\cdots \otimes 
\Gamma^i)_{sym}\ ,\\ 
L_5&\equiv (
\Gamma^5 \otimes {\bf 1}   \otimes {\bf 1} \otimes\cdots \otimes {\bf 1}+ {\bf 1}\otimes 
\Gamma^5 \otimes 
\cdots \otimes {\bf 1} + \cdots + {\bf 1}\otimes {\bf 1}\otimes \cdots \otimes \Gamma^5)_{sym}\ 
,
\end{align}
\end{subequations}
The above would guarantee that the $L_i$ and $L_5$ solve (\ref{S3F-def}a).
$L_i$'s, however, are not defining an $S^3_F$ of definite radius. To obtain a fuzzy three sphere 
of definite radius, we need to restrict $L_i$'s to specific (reducible) representations of
$SO(4)\simeq SU(2)_L\times SU(2)_R$. If we call such a representation ${\cal R}$, i.e.
\be\label{S3F-JJ}
{X^i}/{l}=P_{\cal R} L_i P_{\cal R}\ ,\qquad\quad
{\cal L}_5=P_{\cal R} L_5 P_{\cal R}\ ,
\ee
where $P_{\cal R}$ is the projector restricting $L$'s to ${\cal R}$. Size of the matrices $J$ 
which is defined in  \eqref{S3F-JJ} is \footnote{We would like to emphasize that the ${\cal L}_5$ which 
is used in the Hamiltonian \eqref{Matrix-model-Ham} is only subject to \eqref{S3F-JJ} and 
\eqref{size-rep}; 
eqs. \eqref{calR} and \eqref{size-dim}  only give a specific solution.}
\be\label{size-rep}
J= dim\  {\cal R}\ .
\ee

The requirement that $\sum (X^i)^2=R^2$ is proportional to identity matrix, utilizing Schur's 
Lemma, implies that ${\cal R}$ is a sum of irreducible \rep s of $SO(4)$
moreover, these \rep s should have the same $\sum (L^i)^2$ eigenvalues.
This implies that \cite{sunjay1}
\[
{\cal R}=(\frac{k}{2},\frac{n-k}{2})\oplus (\frac{n-k}{2},\frac{k}{2}) \ ,   
\]
where $(j_L,j_R)$ denotes an irreducible \rep\ of  $Spin(4)=SU(2)_L\times SU(2)_R$ and $j_L$ and 
$j_R$ are integer or half-integer valued.
In addition, in order $X^i$ to be non-trivial in this \rep , $L_i$ must be  
a map between the two irreducible components of ${\cal R}$, which  implies that $k=n-k\pm 1$.
This relation restricts $n$ to be an {\it odd} integer and also
\be\label{calR}
{\cal R}=(\frac{n-1}{4},\frac{n+1}{4})\oplus (\frac{n+1}{4},\frac{n-1}{4}) \ ,   
\ee
Dimension of \rep\ ${\cal R}$, i.e. size of matrices $J$, is
\be\label{size-dim}
J=2\cdot [ 2(\frac{n+1}{4})+1][ 2(\frac{n-1}{4})+1]=\frac{1}{2}(n+1)(n+3), \qquad\ n=odd.
\ee
Therefore, the allowed values for size of the matrices, unlike the $S^2_F$ case, cannot take any 
arbitrary integer values. In particular the lowest $J$'s are $4, 6, 12, \cdots$.

Using the above one can show that \cite{sunjay2}
\be\label{radius-Jn}  
\begin{split}
\sum (X^i/l)^2 &=\sum (P_{\cal R} L^i P_{\cal R})(P_{\cal R} L^i P_{\cal 
R})=\frac{1}{2}(n^2+4n+3)\cr
 &= \left(\frac{R}{l}\right)^2= J\ .
\end{split}
\ee

In sum, the $J\times J$ matrices are filled up with $Spin(4)\simeq SU(2)\times SU(2)$ \rep s
which are of the form ${\cal R}\otimes {\cal R}$ where ${\cal R}$ is given in 
\eqref{calR}, and size of the matrices, $J$ and radius of the $S^3_F$ are related as
given in \eqref{size-dim} and \eqref{radius-Jn}.
As an example, for the lowest $J$, $J=4$,  the above simply reduces to using the $\Gamma$ matrix 
basis for $4\times 4$ matrices, i.e. ${\bf 1},\ \Gamma^5,\ \Gamma^i,\ \Gamma^i\Gamma^5,\ 
\frac{1}{2}\Gamma^{ij}(1\pm \Gamma^5).$


\end{document}